\documentclass[10pt, conference, letterpaper]{IEEEtran}

\usepackage[bookmarks=false]{hyperref}
\usepackage{wrapfig}

\usepackage{mathptmx}       
\usepackage{helvet}         
\usepackage{courier}        
\usepackage{type1cm}        
\usepackage{makeidx}         
\usepackage{graphicx}        
\usepackage{multicol}        
\usepackage[bottom]{footmisc}

\usepackage{subfigure}

\usepackage{cite}

\usepackage{breakurl}

\usepackage{todonotes}
\usepackage{color}

\usepackage{enumitem}

\usepackage{xspace}
\newcommand{\etal}{\textit{et al.}\xspace}

\newcommand{\ie}{\textit{i.e.,}\xspace}
\newcommand{\eg}{\textit{e.g.,}\xspace}

\mathchardef\mhyphen="2D

\newcommand{\framework}{\textsl{\mbox{PRIMA}}\xspace}

\renewcommand{\paragraph}[1]{\medskip \noindent \textbf{#1.\ }}

\begin{document}

\title{\framework: \underline{P}rivacy-P\underline{r}eserving \underline{I}dentity and Access \underline{M}anagement at Internet-Sc\underline{a}le}

\author{
\IEEEauthorblockN{Muhammad~Rizwan~Asghar}
\IEEEauthorblockA{Department of Computer Science}
\IEEEauthorblockA{The University of Auckland }\\
\and
\IEEEauthorblockN{Michael Backes}
\IEEEauthorblockA{CISPA, Saarland University \& MPI-SWS} 
\IEEEauthorblockA{Saarland Informatics Campus} 
\and
\IEEEauthorblockN{Milivoj Simeonovski}
\IEEEauthorblockA{CISPA, Saarland University}
\IEEEauthorblockA{Saarland Informatics Campus}

}

\maketitle

\begin{abstract}

The management of identities on the Internet has evolved from the traditional approach (where each service provider stores and manages identities) to a federated identity management system (where the identity management is delegated to a set of identity providers). 
On the one hand, federated identity ensures usability and provides economic benefits to service providers. 
On the other hand, it poses serious privacy threats to users as well as service providers. 
The current technology, which is prevalently deployed on the Internet, allows identity providers to track the user's behavior across a broad range of services.

In this work, we propose \framework, a universal credential-based authentication system for supporting federated identity management in a privacy-preserving manner. 
Basically, \framework does not require any interaction between service providers and identity providers during the authentication process, thus preventing identity providers to profile users' behavior. 
Moreover, throughout the authentication process, \framework provides a mechanism for controlled disclosure of the users' private information.
We have conducted comprehensive evaluations of the system to show the feasibility of our approach.
Our performance analysis shows that an identity provider can process 1,426 to 3,332 requests per second when the key size is varied from 1024 to 2048-bit, respectively.
\end{abstract}

\section{Introduction}

Authentication is an essential step that aims at accurately identifying the user making a request. 
Traditionally, identities of users were managed separately by each service provider. 
Such a traditional approach is not only impractical for service providers (\ie requiring additional management burden), but is also bad from the users' point of view (\ie they have to manage different credentials for different services).

Federated identity enables users to use the same identity across several services.
Today, many service providers have adopted the concept of federated identity due to its economic advantages and usability benefits.
For service providers, federated identity offers a replacement of standard identity management services; while, for the users, it simplifies the authentication across multiple services, \ie they do not have to remember a pair of user name and password for each service they consume.

One of commonly used federated identity management 
solutions is OpenID~\cite{OpenID:2006} -- an open standard -- adopted by over one million websites for over one billion OpenID enabled user accounts~\cite{OpenIdStat}. 
Typically, the OpenID infrastructure consists of two main entities, an identity provider and a service provider. 
The identity provider is responsible for user registration and management of OpenID accounts, while the service provider offers OpenID-enabled services, which can authenticate users based on their OpenID accounts. 
Using a single OpenID account, users are able to authenticate to multiple service providers. 
After a user is registered with an OpenID provider, she can consume services offered by OpenID-enabled service providers, which can let users authenticate using their OpenID credentials, instead of creating and managing new identities.

For authenticating the requesting user, the service provider forwards the request to the identity provider. 
After the authentication, the identity provider sends a signed confirmation to the user. 
The user forwards this signed confirmation to the service provider, which can finally authenticate the user.
The OpenID Connect \cite{Sakimura:2014:OpenID} -- an extension of OpenID that incorporates OAuth 2.0 \cite{Hardt:2012:OAuth} --  enables authorization besides user authentication. 
More specifically, users choose a piece of personal information a service provider can get access to.

Although the OpenID Connect promises a great deal of usability by offering both authentication and authorization, it puts the privacy of users in danger. 
In all federated identity management solutions (including OpenID, OAuth and OpenID Connect), an identity provider is contacted every time a user visits any service provider, which implies that users' activities are unnecessarily logged by identity providers. 
A curious identity provider could easily profile the users and sell this private data to third-parties, thus invading the privacy of users.

\paragraph{Our Contribution}
In this work, we present \framework, a privacy-preserving federated login, which offers advantages in terms of ensuring privacy.
\framework is deployable and efficient at Internet-scale.
To assist the design of \framework, we propose a modified version of the cryptographic construction presented in Oblivion~\cite{Oblivion:2015}, assuring minimal exposure of private data and scalability.  

In this work, we aim at preserving the privacy of users by modifying the way federated identity management solutions work. 
In the system we propose, an identity provider issues credentials to users after the registration. 
Users store locally and use these credentials when they want to consume services, without requiring any interaction with identity providers. 
Our approach ensures not only users' privacy but also prevents unnecessary interaction between service providers and identity providers. 
Without loss of generality, \framework enables users to have full control over information shared with service providers, \ie allowing users to grant or revoke access at any time. 
Even more, to preserve privacy and to support several real world scenarios, \framework introduces an inference engine for generating attribute specific proofs. 
More specifically, the inference engine generates proofs that prove the possession of a particular attribute without disclosing private information. 
Moreover, a service provider cannot learn more than what is required or permitted by the user. 
The cryptographic technique we use is such that a service provider can verify an arbitrary number of attributes with merely a single verification.

\section{Motivating Scenarios}
\label{sec:motivation}
Let us consider two real world scenarios that require authentication, namely an online bank and an online cinema. 
The first scenario, \ie an online bank, requires a full disclosure of Personally Identifiable Information (PII); whereas, the online cinema has to verify only the age of the user and the geographical location.

\paragraph{Online banking - Full disclosure of PII} 
Consider a user who wants to open a bank account with an online bank. 
Instead of going in person to identify herself, the user wants to use her federated login issued by an authority, \ie identity provider. 
For an opening account request to be lawful, the bank needs to get the full data of the user including her full name and current address.
In such a scenario, the user has to disclose almost all of her private information required by the bank to be able to use the federated login for accessing her bank account. 

\paragraph{Online Cinema - Minimal disclosure}
In our second scenario, we consider a user who wants to access online services that do not require PII, but rather a proof that the user is over a certain age and is not a robot (is a human). 
In order to access the service, instead of disclosing her personal information, to preserve her privacy, she wants to provide only proof that she is over the age required for the selected movie, and that she is registered with some authority.
 
%

\section{Design Overview}
\label{sec:overview}

\begin{figure*} [htp]
	\centering
	\includegraphics[width=.55\textwidth]{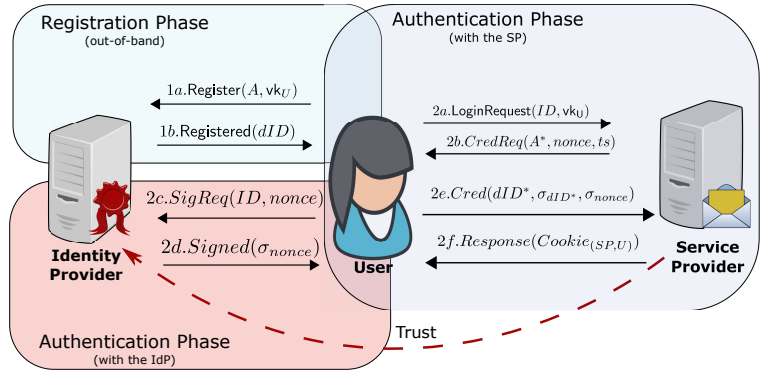}
	\caption{\framework: By providing her attributes (Step 1a), a user gets registered with an Identity Provider (IdP) by receiving signed attributes (Step 1b).
	Later, a user can visit a Service Provider (SP) that trusts the IdP.
	To authenticate to the SP, the user interacts with the SP (Step 2a) and receives a set of requested attributes (Step 2b). 
	To show that her identity is valid, she asks the IdP to sign the nonce, which could be received from the SP (Step 2c). 
	After the validity check, the IdP signs the nonce and sends it back to the user (Step 2d). 
	Finally, she forwards the signed nonce back to the SP along with her digital credential (Step 2e). 
	The SP approves her access (Step 2f).
	}
	\label{fig:general_framework}
	\vspace{-3mm}
\end{figure*}


\subsection{System Entities}
Our design does not require any additional entity, \ie we only rely on entities that are inherently present in the legacy infrastructure, namely an identity provider, a service provider and a user.

\paragraph{Identity Provider (IdP)} 
It is responsible for issuing (or revoking) credentials to the users. 
These credentials are issued after a certain policy of the IdP is fulfilled. 
This policy might include a verification of users' attributes, \ie determining whether the user is associated with the identity, which could be accomplished out-of-band. 

\paragraph{Service Provider (SP)} 
These are entities that allow users access to particular services. 
The approval of the users' access may depend on the certified attributes they receive from the user. 
A trust relation between the SP and the IdP has to be in place for the users to be eligible to use the IdP for accessing the SP.



\paragraph{User (U)} 
On the one hand, users are consumers of the service provided by the SP. 
On the other hand, they are provider of knowledge -- by disclosing their private information to IdPs and SPs.

The basic workflow we consider within this paper is as follows. 
The user contacts the SP to consume some of its services. 
The SP protects its resources and authorizes only legitimate (\ie registered) users. 
To this end, the SP does not deploy its authentication management system but instead relies on a mechanism supported by federated identity. 
To use the federated identity, the user has to register with an IdP (only if she is not already registered), get her attributes certified by the IdP, and finally use them to authenticate with the SP.

\subsection{System Goals}
The main focus of our system is to provide federated authentication while protecting the users' privacy, such that the IdP cannot correlate the authentication request to the destination SP. 
More specifically, the IdP cannot profile the user's online behavior by tracking the authentication requests.

\paragraph{Profile unlinkability}
The IdP should not be able to correlate the authentication requests to the corresponding users. 
In other words, the IdP should not be able to profile the user's behavior.

\paragraph{Selective disclosure}
A service provider should not learn anything beyond the information that is required for providing the service. 
In some cases, this information might be required for eligibility checking, \eg age requirement. 

\paragraph{Non-impersonation} 
The non-impersonation property requires that no one except the user who is in a possession of a secret can be authorized. Evermore, the IdP should also not be able to impersonate its users.

\paragraph{Deployability}
The system is required to be deployable from a technical
perspective. 
Given the extensive deployed federated identity solutions (\ie OpenID and SAML), the systems should be incrementally deployable as a complement to the existing infrastructure.

\paragraph{Non-Goals}
We do not aim at providing anonymous credentials like \cite{anoncredentials}. 
Our IdP is trusted and has a full knowledge of the users' PII. 
However, our system supports ``weak'' variant of anonymous credentials. 
The IdP can issue an authentication token for the requesting user without disclosing any of her PII. 
For instance, if the service provider requests a proof that the consumer of its services is located in a particular country, then the IdP can be instructed to generate a valid authentication token disclosing only one a single attribute which is the country of residence.

%

\subsection{Threat Model}
We consider the threat model as considered in the OpenID protocol~\cite{lodderstedt:2013:oauth}.
We aim to achieve aforementioned goals without changing the current infrastructure. 
Our adversary has a full access to the network resources between the communicating parties and he can eavesdrop any of the communicated data.


We consider the IdP to be honest-but-curious, meaning it is honest to issue or revoke credentials but curious to profile the users' behavior. 
We do not fully trust the SP. 
We assume a malicious SP that (1) wants to benefit from the user's data and (2) reuse the user's attributes or session cookies to authenticate. 
Finally, we assume that a user space is not compromized.

\subsection{Key Idea}
The key idea is to use \emph{credential-based authentication} without direct communication and data exchange between the IdP and the SP during the authentication process. 
The IdP issues a unique digital assertion that carries digital information, \ie attributes about the user's identity. 
To achieve the minimal disclosure requirement defined in the system goals, our system relies on the cryptographic construction for controlled disclosure of the attributes proposed in Oblivion~\cite{Oblivion:2015}. 
Using this construction, the authenticating user can prove possession of the identity by disclosing a minimal set of required attributes. 
Since the attributes are issued without specifying a particular service and can be verified without interaction with the IdP, all the requests are undetectable for the IdP.
We define the digital credential $dID$ issued by the IdP to be a tuple of the form.

\begin{equation}
dID = \left\langle A, \sigma_{A}, vk_U, T \right\rangle
\label{eq:dID}
\end{equation}

where $A = \{a_1, a_2,\ldots,a_n\}$ is a list of the user's attributes (where an attribute could be a user name or her address), $\sigma_{A} = \{ \sigma_{a_1},\sigma_{a_2},\ldots,\sigma_{a_n} \} $ is a list of signed attributes, $vk_U$ is the verification key for verifying messages signed by the user, and $T=\{t_{isu},t_{exp}\}$ is a set of time values that define the issuing ($t_{isu}$) and the expiration date ($t_{exp}$).

\section{Communication Protocol}
\label{sec:approach}
Figure~\ref{fig:general_framework} presents the proposed architecture to achieve the aforementioned goals. 
Our system consists of two phases, namely registration and authentication. 
The registration phase is generally a one-time step when the user signs up (\ie opens an account) with the IdP. 
This phase could also be accomplished out-of-band. 
During the authentication phase, the user proves her eligibility to use the service provided by the SP. 
In the following, we describe the sequence of events that the user has to follow in order to authenticate with the SP.

\begin{figure*} [htp]
	\centering
	\includegraphics[width=.55\textwidth]{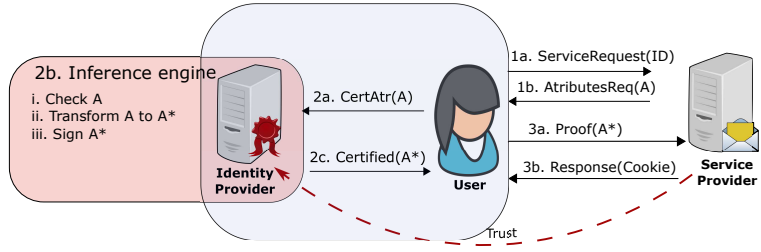}
	\caption{Inference engine: To request a service from the SP, the user interacts with the SP (Step 1a) and receives a set of required attributes (Step 1b). 
	To protect her privacy, she asks the IdP to transform the attributes into logical proofs such that the selected criteria are fulfilled (Step 2a). 
	The IdP checks the attributes, transforms them into logical proofs, signs them accordingly (Step 2b), and sends them back to the user (Step 2c). 
	Finally, she forwards the signed attributes back to the SP along with her digital credential (Step 3a). 
	The SP approves her access (Step 3b).}
		\vspace{-3mm}
	\label{fig:inference}
\end{figure*}

\begin{itemize}

	\item ($U \rightarrow IdP$): 
	Figure~\ref{fig:general_framework} Step 1a \\
The user U registers with the identity provider IdP by providing a set of her ($n$) attributes $A = \{a_1, a_2, \ldots, a_n\}$ along with her verification key $\ensuremath{\mathsf{vk}}_U$. 
Each attribute $a_i\in A$ is a pair \textsf{$\langle$KEY, VALUE$\rangle$}, representing name of the attribute and value specific to each user, respectively.

	\item ($IdP \rightarrow U$): 
	Figure~\ref{fig:general_framework} Step 1b \\
Upon a successful verification of the provided attributes, the IdP issues a list of ($n$) signed attributes $\sigma_A = \{\sigma_{a_1}, \sigma_{a_2}, \ldots, \sigma_{a_n}\}$ and sends it back in the form of a digital credential (dID) to the requesting user. 
We employ the attribute signing scheme presented in Oblivion~\cite{Oblivion:2015} and bind every attribute with the verification key of the user and an expiration date. 
More formally, for each attribute $a_i \in A$, the IdP computes:
\[\sigma_{a_i} = Sign(\ensuremath{\mathsf{sk}}_{IdP}, a_i || \ensuremath{\mathsf{vk}}_U || t_{exp})
\]
where $Sign$ is a signing algorithm and $||$ denotes concatenation.

	\item ($U \rightarrow SP$): 
	Figure~\ref{fig:general_framework} Step 2a \\
Once the user is in possession of authentication credentials, she can send a request for consuming a service to an SP that supports credential-based authentication. 
The user requests access providing her unique ID (\ie her verification key $\ensuremath{\mathsf{vk}}_U$).

	\item ($SP \rightarrow U$): 
	Figure~\ref{fig:general_framework} Step 2b \\
The SP responds with a credential request and a fresh nonce, requesting a set of user's attributes $A^*$. 
Depending on the service that the SP offers, the set of requested attributes could vary from a simple unique identification to fine-grained attributes, such as date of birth, location and biometric attributes.

\item ($U \rightarrow IdP$): 
	Figure~\ref{fig:general_framework} Step 2c \\
To prove that the user's account is actual (\ie neither blocked or revoked), the user needs to get a signed nonce from the IdP. 
One the nonce is received from the SP, she contacts the IdP on the fly to request a signature on the nonce. 

\item ($IdP \rightarrow U$): 
	Figure~\ref{fig:general_framework} Step 2d \\
The IdP checks the status of the user account, and once verified that the account is valid, it signs the nonce and sends it back to the user.

	\item ($U \rightarrow SP$): 
Figure~\ref{fig:general_framework} Step 2e \\
Finally, upon receiving the signed nonce, the user sends to the SP her digital credential,
$dID^* = \left\langle A^*, P_{\sigma_{A^*}}, vk_U, TS, sID \right\rangle$, where $P_{\sigma_{A^*}}$ is a packed signature of (requested) attributes $A^*$, along with a signature $\sigma_{dID^*}$ over the $dID^*$ and the signed nonce.
For preventing replay attacks, a timestamp $TS$ and a session sID is included in the message. 


	\item ($SP \rightarrow U$): 
Figure~\ref{fig:general_framework} Step 2f \\
Upon receiving the credential $dID^*$, the SP verifies it before authorizing the user. 
First, it verifies the signature $\sigma_{dID^*}$ over the $dID^*$. 
Next, it checks if the authentication session and time values are correct. 
Finally, the SP verifies the attributes and the validity of the packed signature. 
After a successful verification, the SP issues an authorization token for the requesting user. 

\end{itemize}

\begin{figure*}[htp]
	\centering
	\subfigure[]{
		\includegraphics[width=.3\textwidth]{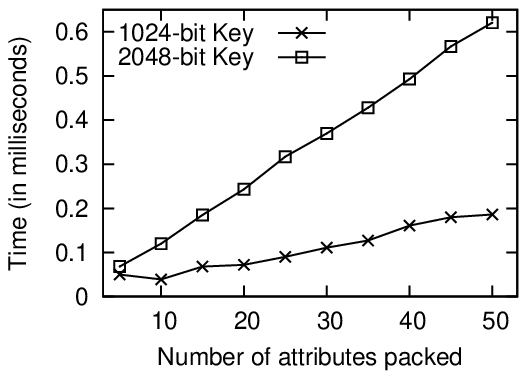}
		\label{fig:attribute-pack}
	}
	\subfigure[]{
		\includegraphics[width=.3\textwidth]{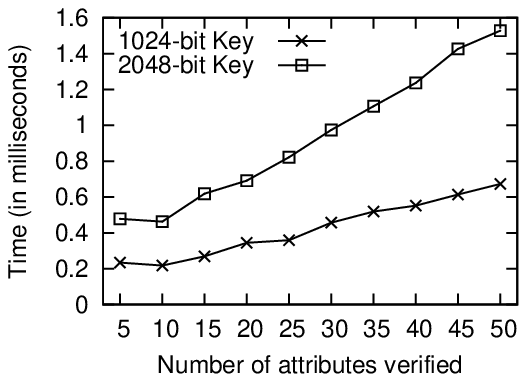}
		\label{fig:attribute-verify}
	}
	\subfigure[]{
		\includegraphics[width=.3\textwidth]{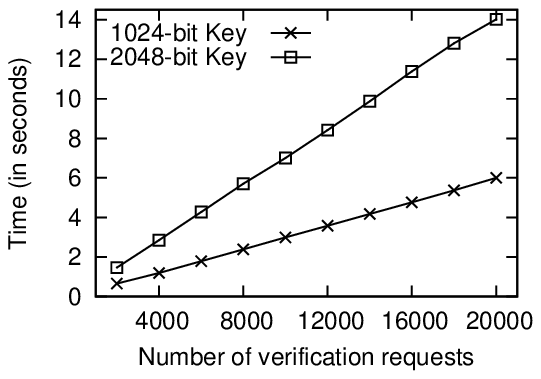}
		\label{fig:verification-requests}
	} 	
	\caption{Performance overhead of \subref{fig:attribute-pack} packing attributes, \subref{fig:attribute-verify} verifying attributing and \subref{fig:verification-requests} verifying user requests.}
\label{fig:req-verify}
	\vspace{-3mm}
\end{figure*}

Upon a successful completion of the authentication phase, the user is authorized by the SP and receives an access token or a cookie. 
The expiration policy of the credentials whether they are in a form of access token or a cookie are out of the scope of this work. 
Every time the user requests access to the SP, she has to go over the authentication steps.

\section{Inference Engine}
\label{sec:inference}
In order to support real world scenarios presented in Section~\ref{sec:motivation} (say the second scenario), and preserve privacy of the users, we introduce a new module named \emph{inference engine}.
This module is in charge of transforming the real users' attributes into a logical proof that shows possessing of the attributes without disclosing their real value.
To explain the importance of such a module, in the following, we present the simple scenario from the Section \ref{sec:motivation} from the perspective of the inference engine (Figure \ref{fig:inference}).

The user, say Alice, visits online cinema (say, OnlineKino) in order to watch a movie. 
OnlineKino asks Alice to prove that she is over 16 and to provide her country of residence in order to be able to watch the selected movie.
Alice receives a list of all required attributes from OnlineKino along with a nonce (in this case the age attribute, and country).
Alice chooses to use her IdP and would provide authorization to OnlineKino to use the required set of attributes.
After Alice is authenticated with her IdP and after the IdP of Alice receives authorization, Alice decides to give temporary access to the following attributes:
\begin{itemize}
	\item Proof that Alice is over 16
	\item Country
\end{itemize}

On her confirmation, the IdP generates an ID token bound with a nonce (initially generated by the SP). 
Finally, Alice sends the token back to the SP. 
After the verification by the SP, she gets access to the requested movie.
Note that the communication between the IdP and the SP is made through Alice.

Figure~\ref{fig:inference} presents a sequence of events that the protocol has to follow in order to answer the request from the users for issuing the requested proof.

\begin{itemize}
	
	\item ($U \rightarrow SP$): 
	Figure~\ref{fig:inference} Step 1a \\
The user sends her request to the SP for consuming a service. 
Same as in Figure~\ref{fig:general_framework}, the user sends her ID along with the request.

	\item ($SP \rightarrow U$): 
	Figure~\ref{fig:inference} Step 1b \\
The SP responds with an attributes request message, requesting a set of user's attributes.

	\item ($U \rightarrow IdP$): 
	Figure~\ref{fig:inference} Step 2a \\
The user asks her IdP to certify the set of attributes $A$. 
If Alice is willing to blind some of the attributes to only show that a condition is met, such as certain age, she requests this transformation from the IdP.   

	\item ($IdP$): 
	Figure~\ref{fig:inference} Step 2b \\
	The IdP activates its inference module to generate a proof for the user's attributes.
	\begin{enumerate}[label=(\roman*)]

		\item Initially, the module checks the existence of the attributes.

		\item When privacy of the attributes is requested by the user, the module transforms the attributes from their original form into a logical proof that the selected criteria are fulfilled. 
		For example, if the user is requesting a proof that she is over a certain age without disclosing the current age, the module generates a statement (\eg \emph{Alice is over 16}).

		\item Finally, such transformation in a form of a statement is signed by the IdP.

	\end{enumerate}

\item ($IdP\rightarrow U$): 
	Figure~\ref{fig:inference} Step 2c \\
The IdP sends the certified attributes to the user.

\item ($U\rightarrow SP$): 
	Figure~\ref{fig:inference} Step 3a \\
The user receives the request and responds with a packed signature of attributes $P_{\sigma_{A^*}}$, where $A^*$ is the set of the requested attributes. 
Finally, the user forwards her digital credential along with the certified attributes to the SP.

\end{itemize}

\section{Performance Analysis}
\label{sec:performance}

\begin{figure} [htp]
\centering
\includegraphics[width=.3\textwidth]{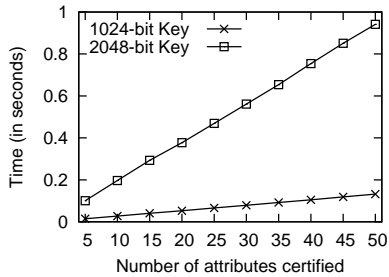}
\caption{Performance overhead of certifying attributes.}
\label{fig:attribute-gen}
\vspace{-3mm}
\end{figure}

We provide an experimental demo implementation and subsequently evaluate the performance overhead for all the entities in the system. 
The implementation prototype is written in Java and run on a standard notebook with 2.7 GHz Intel core processor and 8 GB RAM running Windows 7 Enterprise (64-bit). 

\paragraph{IdP: Certifying User Attributes} 
One part of the registration phase is the certification, \ie signing of the users' attributes. 
Figure~\ref{fig:attribute-gen} illustrates the computational overhead for the signing algorithm. 
For the experiment, we generated up to 50 attributes with varying size of the signing key. 
On average, the IdP takes 2.64 and 18.92 milliseconds (ms) per attribute for 1024 and 2048-bit keys, respectively.
The certification time grows linearly in the number of attributes. 
The registration phase usually is a one-time activity per user; therefore, this computational overhead is acceptable.

\paragraph{U: Packing Attributes} 
Figure~\ref{fig:attribute-pack} shows the computational overhead of packing attributes on the user's side. 
We observe that the time taken by the packing algorithm increases linearly with increase in the number of attributes, with an additional overhead for larger key sizes. 
For 50 attributes, it took between 0.19 ms (for a 1024-bit key) and 0.62 ms (for a 2048-bit key).

\paragraph{SP: Verifying Attributes and Requests} 
Next, we measured the rate at which an IdP can verify attributes. 
Figure~\ref{fig:attribute-verify} shows that for verifying 50 attributes, the SP takes 0.67 ms and 1.53 ms for 1024-bit and 2048-bit keys, respectively. 

Finally, Figure~\ref{fig:verification-requests} shows the efficiency of the authentication phase. 
Assuming that every authentication request requires
20 attributes\footnote{We assume a worst-case scenario with a disclosure of 20 private attributes.}, and gradually increasing the number of user requests from 2,000 to 20,000, we observed an essentially linearly-growing overhead. 
On average, an IdP can process 3,332 and 1,426 requests per second for 1024-bit and 2048-bit keys, respectively.

\section{Discussion}
\label{sec:discussion}
It is known that the growing number of online services make the traditional approach for authentication and identity management infeasible.
To solve this issue, federated identity is a desirable feature, not only for users but also for service providers.
However, we believe that the privacy aspect has yet to be considered when designing new or upgrading current identity management solutions. 

In this work, we proposed a universal credential-based authentication system for supporting federated identity management in a privacy-preserving and efficient manner. 
Our approach prevents identity providers from profiling the users' behavior and also gives users full control over their private attributes shared with service providers. 
Finally, we showed that the system we propose is efficient and scalable, and can be deployed today.

An alternative solution to achieve privacy in the currently deployed infrastructure is to modify the traditional identity and access management protocols to ensure unlinkability between the service provider and the identity provider.
To this end, we suggest that users should receive the request from the service provider.
Before forwarding this request to the identity provider, the user should strip off the service provider specific information from the request.
After receiving the response from the identity provider, the user can forward it to the service provider.
Without loss of generality, the identity provider can use the inference engine to generate request proofs.

\section{Related Work}
\label{sec:related}
There are a number of initiatives in the EU concerning privacy in identity management \cite{Camenisch-DIM05-Privacy, Ardagna-JCS10-PRIME}, where pseudo-identities \cite{Chaum-CACM81-Pseudonyms} and anonymous credentials \cite{Camenisch-CCS02-Idemix} are the underlying techniques.
On one hand, anonymous credentials allow users to perform offline authentication, \ie without requiring interaction with the identity provider.
On the other hand, it is challenging to do revocation due to offline nature of issued credentials.
Unfortunately, anonymous credentials face serious scalability issues when it comes to revocation.

Bertino \etal \cite{Bertino-DEB09-Privacy} present an approach to the verify digital identities for cloud platforms.
However, they do leave unlinkability and delegatable authentication problems as open.
Both issues have been addressed by Chow \etal in \cite{Chow-ACNS12-SPICE}.
However, we argue that the scheme used in \cite{Chow-ACNS12-SPICE} is based on Identity-Based Encryption (IBE) \cite{Waters-EUROCRYPT05-IBE} that presents challenges when it comes to revocation.

PseudoId~\cite{Dey:2010:PseudoID} is a prototype of a protocol that protects users from disclosure of private login data possessed by the identity providers. 
It utilizes blind digital signatures to protect user's real identity. 
While this protocol is designed to be one-way, and unlinkable federated login system, it does not support selective disclosure of private data neither linking between the users' information and their identities, thus limiting its usability to only limited scenarios. 

Ardagna \etal~\cite{Ardagna:2010} propose an extension of XACML and SAML to achieve privacy-preserving access control. 
Similar to our solution, they assume that the identity provider has a complete knowledge of users' attributes and the requester can selectively disclose his attributes. 
Unlike our protocol, they use anonymous credentials for protecting the users' privacy.

In order to overcome the privacy issues with respect to the cloud providers, Nunez \etal integrate proxy re-encryption into OpenId and SAML~\cite{Nunez:2012:OpenIDproxy,Nunez:2014:BlindIdM}. 
The latter one, BlindIdM is a privacy-preserving realization for IDaaS (Identity Management as a Service).
In their proposal, they do not fully outsource the identity management system, but instead they keep the authentication at the host organization that encrypts the users' identity information before outsourcing to the identity provider.

\section{Conclusions and Future Directions}
\label{sec:conclusion}
We proposed \framework, a universal credential-based authentication system for supporting federated identity management in a privacy-preserving manner.
\framework does not require any interaction between the identity provider and the service provider, thus preventing active profiling of the users. 
Moreover, throughout the authentication process, \framework provides a mechanism for controlled disclosure of the users' private information and mechanism for building logical proofs for certain tributes without disclosing their real value.
Finally, we conducted comprehensive evaluations of the system to show the feasibility of our approach.
Our performance analysis showed that an identity provider can process 1,426 to 3,332 requests per second when the key size varies from 1024 to 2048-bit, respectively.

As future work, we plan to make a user study about the usability of the proposed system and formalize the corresponding security and privacy definitions.

\bibliographystyle{IEEEtran}
\bibliography{references}

\begin{thebibliography}{10}
\providecommand{\url}[1]{#1}
\csname url@samestyle\endcsname
\providecommand{\newblock}{\relax}
\providecommand{\bibinfo}[2]{#2}
\providecommand{\BIBentrySTDinterwordspacing}{\spaceskip=0pt\relax}
\providecommand{\BIBentryALTinterwordstretchfactor}{4}
\providecommand{\BIBentryALTinterwordspacing}{\spaceskip=\fontdimen2\font plus
\BIBentryALTinterwordstretchfactor\fontdimen3\font minus
  \fontdimen4\font\relax}
\providecommand{\BIBforeignlanguage}[2]{{%
\expandafter\ifx\csname l@#1\endcsname\relax
\typeout{** WARNING: IEEEtran.bst: No hyphenation pattern has been}%
\typeout{** loaded for the language `#1'. Using the pattern for}%
\typeout{** the default language instead.}%
\else
\language=\csname l@#1\endcsname
\fi
#2}}
\providecommand{\BIBdecl}{\relax}
\BIBdecl

\bibitem{OpenID:2006}
D.~Recordon and D.~Reed, ``{OpenID} 2.0: a platform for user-centric identity
  management,'' in \emph{Proceedings of the 2006 Workshop on Digital Identity
  Management, Alexandria, VA, USA, November 3, 2006}, 2006, pp. 11--16.

\bibitem{OpenIdStat}
``{OpenID} usage statistics,''
  \url{http://trends.builtwith.com/docinfo/OpenID}, 2016, last accessed:
  October 14, 2016.

\bibitem{Sakimura:2014:OpenID}
N.~Sakimura, J.~Bradley, M.~Jones, B.~de~Medeiros, and C.~Mortimore, ``{OpenID
  Connect} core 1.0,'' \emph{The OpenID Foundation}, 2014.

\bibitem{Hardt:2012:OAuth}
D.~Hardt, ``The {OAuth} 2.0 authorization framework,'' 2012, rfc 6749.

\bibitem{Oblivion:2015}
M.~Simeonovski, F.~Bendun, M.~R. Asghar, M.~Backes, N.~Marnau, and P.~Druschel,
  ``Oblivion: Mitigating privacy leaks by controlling the discoverability of
  online information,'' in \emph{{ACNS} 2015}.

\bibitem{anoncredentials}
J.~Camenisch and A.~Lysyanskaya, ``An efficient system for non-transferable
  anonymous credentials with optional anonymity revocation,'' in \emph{Advances
  in Cryptology - {EUROCRYPT} 2001, International Conference on the Theory and
  Application of Cryptographic Techniques, Innsbruck, Austria, May 6-10, 2001,
  Proceeding}, 2001, pp. 93--118.

\bibitem{lodderstedt:2013:oauth}
T.~Lodderstedt, M.~McGloin, and P.~Hunt, ``{OAuth} 2.0 threat model and
  security considerations,'' 2013.

\bibitem{Camenisch-DIM05-Privacy}
J.~Camenisch, D.~Sommer, S.~Fischer-H{\"u}bner, M.~Hansen, H.~Krasemann,
  G.~Lacoste, R.~Leenes, J.~Tseng \emph{et~al.}, ``Privacy and identity
  management for everyone,'' in \emph{Proceedings of the 2005 workshop on
  Digital identity management}.\hskip 1em plus 0.5em minus 0.4em\relax ACM,
  2005, pp. 20--27.

\bibitem{Ardagna-JCS10-PRIME}
C.~A. Ardagna, J.~Camenisch, M.~Kohlweiss, R.~Leenes, G.~Neven, B.~Priem,
  P.~Samarati, D.~Sommer, and M.~Verdicchio, ``Exploiting cryptography for
  privacy-enhanced access control: A result of the {PRIME} project,''
  \emph{Journal of Computer Security}, vol.~18, no.~1, pp. 123--160, 2010.

\bibitem{Chaum-CACM81-Pseudonyms}
D.~L. Chaum, ``Untraceable electronic mail, return addresses, and digital
  pseudonyms,'' \emph{Communications of the ACM}, vol.~24, no.~2, pp. 84--90,
  1981.

\bibitem{Camenisch-CCS02-Idemix}
J.~Camenisch and E.~Van~Herreweghen, ``Design and implementation of the idemix
  anonymous credential system,'' in \emph{Proceedings of the 9th ACM Conference
  on Computer and Communications Security}, ser. CCS '02.\hskip 1em plus 0.5em
  minus 0.4em\relax New York, NY, USA: ACM, 2002, pp. 21--30.

\bibitem{Bertino-DEB09-Privacy}
E.~Bertino, F.~Paci, R.~Ferrini, and N.~Shang, ``Privacy-preserving digital
  identity management for cloud computing.'' \emph{IEEE Data Eng. Bull.},
  vol.~32, no.~1, pp. 21--27, 2009.

\bibitem{Chow-ACNS12-SPICE}
S.~S.~M. Chow, Y.-J. He, L.~C.~K. Hui, and S.~M. Yiu, \emph{{SPICE} -- Simple
  Privacy-Preserving Identity-Management for Cloud Environment}.\hskip 1em plus
  0.5em minus 0.4em\relax Berlin, Heidelberg: Springer Berlin Heidelberg, 2012,
  pp. 526--543.

\bibitem{Waters-EUROCRYPT05-IBE}
B.~Waters, \emph{Efficient Identity-Based Encryption Without Random
  Oracles}.\hskip 1em plus 0.5em minus 0.4em\relax Berlin, Heidelberg: Springer
  Berlin Heidelberg, 2005, pp. 114--127.

\bibitem{Dey:2010:PseudoID}
\BIBentryALTinterwordspacing
A.~Dey and S.~Weis, ``{PseudoID}: Enhancing privacy in federated login,'' in
  \emph{Hot Topics in Privacy Enhancing Technologies}, 2010, pp. 95--107.
  [Online]. Available: \url{http://www.pseudoid.net}
\BIBentrySTDinterwordspacing

\bibitem{Ardagna:2010}
C.~A. Ardagna, S.~D.~C. di~Vimercati, G.~Neven, S.~Paraboschi, F.~Preiss,
  P.~Samarati, and M.~Verdicchio, ``Enabling privacy-preserving
  credential-based access control with {XACML} and {SAML},'' in \emph{10th
  {IEEE} International Conference on Computer and Information Technology, {CIT}
  2010, Bradford, West Yorkshire, UK, June 29-July 1, 2010}, 2010, pp.
  1090--1095.

\bibitem{Nunez:2012:OpenIDproxy}
D.~Nu{\~{n}}ez, I.~Agudo, and J.~Lopez, ``Integrating {OpenID} with proxy
  re-encryption to enhance privacy in cloud-based identity services,'' in
  \emph{4th {IEEE} International Conference on Cloud Computing Technology and
  Science Proceedings, CloudCom 2012, Taipei, Taiwan, December 3-6, 2012},
  2012, pp. 241--248.

\bibitem{Nunez:2014:BlindIdM}
D.~Nu{\~{n}}ez and I.~Agudo, ``{BlindIdM: A} privacy-preserving approach for
  identity management as a service,'' \emph{Int. J. Inf. Sec.}, vol.~13, no.~2,
  pp. 199--215, 2014.

\end{thebibliography}


\end{document}